\documentclass[prl,twocolumn,aps,superscriptaddress,nofootinbib]{revtex4-1}

\usepackage{amsmath}
\usepackage{esint}   
\usepackage{graphicx}
\usepackage{hyperref}
\usepackage{color}
\usepackage{xcolor}
\usepackage{enumerate}
\usepackage[utf8]{inputenc} 
\advance\voffset -0.2in

\newcommand{\be}{\begin{eqnarray}}
\newcommand{\ee}{\end{eqnarray}}

\newcommand{\ave}[1]{\left\langle #1 \right\rangle}
\newcommand{\avet}[1]{\langle #1 \rangle}  

\newcommand{\gev}{{\rm \, GeV}}

\newcommand{\var}{\mathrm{Var}}

\newcommand{\nrun}{n_{\rm run}}
\newcommand{\neve}{n_{\rm events}}
\newcommand{\neveN}{144393}

\definecolor{vkcolor2}{HTML}{336600}

\newcommand{\aba}[1]{{\color{red} \Large X}}  

\newcommand{\overbar}[1]{\mkern 1.5mu\overline{\mkern-1.5mu#1\mkern-1.5mu}\mkern 1.5mu}
\newcommand{\nb}{\overbar{N}}

\begin{document}

\title{Mapping the QCD phase diagram with statistics friendly distributions}

\author{Adam Bzdak}
\email{bzdak@fis.agh.edu.pl}
\affiliation{AGH University of Science and Technology,
Faculty of Physics and Applied Computer Science,
30-059 Krak\'ow, Poland}

\author{Volker Koch}
\email{vkoch@lbl.gov}
\affiliation{Nuclear Science Division, 
Lawrence Berkeley National Laboratory, 
Berkeley, CA, 94720, USA}

\begin{abstract}
We demonstrate that the multiplicity distribution of a system located in the vicinity of a first-order 
phase transition can be successfully measured in terms of its factorial cumulants with a  surprisingly
small number of events. This finding has direct implications for the experimental search of a QCD
phase transition conjectured to be located in the high baryon density region of the QCD phase diagram.  
\end{abstract}

\maketitle

One of the key questions of the physics of strong interactions is the possible existence of a first-order 
phase transition accompanied by a critical point. While lattice QCD has established that the
transition at vanishing net-baryon density is an analytic cross over \cite{Aoki:2006we}, the presence
of a first-order transition accompanied by a critical point has been conjectured based on many model
calculations (see e.g. \cite{Stephanov:2004wx,Bzdak:2019pkr} for a review). To search for such a possible
transition in experiment, fluctuations of conserved charges in relativistic heavy ion collisions have
been considered as  promising probes \cite{Jeon:2000wg,Asakawa:2000wh,Stephanov:2008qz,Skokov:2010uh,Stephanov:2011pb,Luo:2011rg,
Luo:2017faz,Herold:2016uvv,Zhou:2012ay,Wang:2012jr,Karsch:2011gg,Schaefer:2011ex,Chen:2011am,
Fu:2009wy,Cheng:2008zh}. Special attention has been paid to the cumulants of the net-baryon or
net-proton\footnote{Experimentally, one is usually restricted to the measurement of cumulants of the net-proton distribution \cite{Aggarwal:2010wy,Adamczyk:2013dal,Rustamov:2017lio} since neutrons are difficult to measure. However, as shown in \cite{Kitazawa:2011wh,Kitazawa:2012at} given fast isospin-exchange processes due to the abundance of pions the connection to the net-baryon number cumulants can be made.} number distribution as they are particularly sensitive to the details of the transition from hadron gas to
quark-gluon plasma in the cross-over region \cite{Skokov:2010uh,Karsch:2011gg} as well as near a
potential critical point \cite{Stephanov:2008qz}. This sensitivity is expected to increase with
the order of the cumulant \cite{Stephanov:2008qz}, the measurement of which is commonly believed to require increasing statistics. 

In this paper we show, quite generally, that it requires surprisingly few events to determine
if a system is located close to a first-order phase transition. This finding has direct implications
on the search for the QCD phase transition, but will also be relevant for any other (mesoscopic)
systems where fluctuation measurements are meaningful. It is well known that the multiplicity
distribution of a system close to a first-order phase transition is a two-component or bi-modal
distribution reflecting the two (dense and dilute) phases. If the system is right at
the transition it has two maxima of equal magnitude, reflecting the equal probability of the two
phases. As one moves away from the transition, one of the maxima becomes smaller, reflecting the fact that away from the transition one phase is much more probable than
the other. Thus, for small systems and not too far from the transition, the presence of the other
phase still shows up in the multiplicity distribution (for a detailed discussion, see
\cite{Bzdak:2018uhv}). As discussed in \cite{Bzdak:2018uhv}, such a two-component multiplicity
distribution, even in the case when one of the components is rather small, has a very characteristic
behavior of its factorial cumulants: with increasing order they increase rapidly in magnitude with
alternating sign (in contrast, ultrarelativistic quantum molecular dymamics (UrQMD) calculations give higher order factorial cumulants consistent with zero \cite{He:2017zpg}). This characteristic may be used to establish the existence of a two-component
multiplicity distribution, which in turn would provide strong evidence that the system is close to a
first-order phase transition.\footnote{A two-component
    distribution could in principle also result from different effects, such as production
    vs. stopping of protons, problems with centrality determination, deuteron enhancement in some
    events, possible issues with a detector etc. However, it seems these effects should be also
    visible at, say, $19.6$ GeV, where the higher order factorial cumulants are consistent with zero
    \cite{Bzdak:2016sxg} and a two-component distribution is not visible.}

Such a characterization requires factorial cumulants of many orders that are commonly
believed to require large statistics. However, as we show, 
the two-component distributions relevant for a first-order phase transition are remarkably
statistics friendly in the sense that for a given and rather limited number of events factorial
cumulants can be reliably extracted to a surprisingly high order. Surprisingly, this is even the
case if the second mode (component) is rather tiny and is difficult to see directly in the multiplicity distribution. This finding, therefore, demonstrates that a search for a
first-order phase transition via fluctuation measurements is practically feasible and does not
require unrealistic levels of statistics.

In the following we illustrate our findings in the context of preliminary results of the STAR
Collaboration. However, our arguments are quite general and are not restricted to
the QCD phase transition.  The preliminary results from the STAR Collaboration for the ratio of
fourth-order over second-order (net)-proton cumulants show an intriguing pattern \cite{Luo:2015ewa}. It
grows rapidly with decreasing beam energy from $\sqrt{s}=19.6$ GeV reaching a large value at $7.7$
GeV. It was argued \cite{Bzdak:2016sxg} that this behavior is caused by a strong increase of
multi-proton correlations with decreasing energy. In addition it was found \cite{Bzdak:2018uhv},
that at the lowest energy,  $\sqrt{s}=7.7\gev$, where the deviation of the cumulants from a Poisson
baseline (or rather binomial due to baryon conservation)  are the largest, the first four (factorial) cumulants, so far measured by STAR, are consistent with a
two-component proton multiplicity distribution, albeit with the second component being rather small. Of course the first four cumulants are not enough to
sufficiently constrain the multiplicity distribution. Therefore, it is essential to measure
(factorial) cumulants of higher order to either confirm or rule out that the underlying distribution
is indeed a two-component one consistent with a first-order phase transition. As we show  this
is possible due to the ``statistics friendly'' properties of these two-component distributions even
for the very limited statistics of the present STAR data set.

Specifically, in this paper we study various proton multiplicity distributions to
evaluate the statistical errors of higher order factorial cumulants. In our studies we choose a
rather small number of events, approximately 
150000 (144393 to be more precise \cite{XL-private}), which is the statistics underlying the STAR measurement for
the most central Au+Au collisions at $\sqrt{s}=7.7$ GeV at RHIC
\cite{Adamczyk:2013dal,Luo:2015ewa}. We will only consider multiplicity distributions
of one species of particles, which are protons in our case.\footnote{It would be
interesting to explore if similar statistics friendly distributions also exist for more than one
species, such as net-proton distributions which involve protons and anti-protons.}

To evaluate the statistical errors numerically, we sample the number of protons, $N$, $\neve=
144393$ times from a given multiplicity distribution $P(N)$. We then
calculate the cumulants, $K_{n}$ and the factorial cumulants $C_{n}$ for $n=2,...,9$.\footnote{As a reminder, the cumulants and factorial cumulants are obtained from the
multiplicity distribution $P(N)$ as
$ K_{n}=\left. \frac{d^{n}}{dt^{n}}\ln \left[\sum_{N}P(N)e^{N t}\right]\right|_{t=0} $,
$ C_{n}=\left. \frac{d^{n}}{dz^{n}}\ln \left[\sum_{N}P(N)z^{N}\right]\right|_{z=1}  $.}
Next we repeat this sampling $\nrun$ times, where $\nrun$ is sufficiently large so that the results
presented below do not depend on $\nrun$. This procedure
then gives us $\nrun$ ``measurements''  or samples of $K_{n}$ and $C_{n}$, 
$\{K^{(1)}_{n},\ldots,K^{(\nrun)}_{n}\}$ and  $\{C^{(1)}_{n},\ldots,C^{(\nrun)}_{n}\}$. 

From these samples we calculate the variance, for example, in the case of
the factorial cumulants $C_{n}$ we have 
\begin{equation}
  \mathrm{Var}\left( C_{n}\right) =\frac{1}{n_{\mathrm{run}}}\sum_{i=1}^{n_{%
  \mathrm{run}}}\left( C_{n}^{(i)}\right) ^{2}-\left( \frac{1}{n_{\mathrm{run}}%
  }\sum_{i=1}^{n_{\mathrm{run}}}C_{n}^{(i)}\right) ^{2}.
\end{equation}
The expected \emph{absolute}
error, $\Delta C_{n}$, is then given by $\Delta C_{n}=\sqrt{\mathrm{Var}\left( C_{n}\right) }$, 
whereas the \emph{relative} error is 
$\Delta C_{n}/C_{n}$, where $C_{n}$ denotes the  true value
directly calculated from the multiplicity distribution $P(N)$. 

An alternative way to calculate the expected error is by means of the delta method 
(see, e.g., \cite{davison2003statistical,Luo:2014rea,Luo:2017faz} for details). In
the case at hand, where we want to calculate the errors for (factorial) cumulants, application of
the delta method is straightforward. Let us discuss this in more detail for the case of the
factorial cumulant.  The
random variables are the moments about zero, $\mu_{k}=\ave{N^{k}}$. Therefore, we express the
factorial cumulant, $C_{n}$, in terms of the moments, $C_{n}=F(\mu_{1}, \ldots, \mu_{n})$. Then
according to the delta method the variance of $C_{n}$ for a sample with $\neve$ events is given by 
\begin{eqnarray}
  \mathrm{Var}\left( C_{n}\right)  &=&\sum_{i=1}^{n}\sum_{j=1}^{n}\frac{%
  \partial F}{\partial \mu _{i}}\frac{\partial F}{\partial \mu _{j}}\mathrm{Cov%
  }\left( \mu _{i},\mu _{j}\right) ,  \label{eq:delta} \\
  \mathrm{Cov}(\mu _{i},\mu _{j}) &=&\frac{1}{n_{\mathrm{events}}}\left( \mu
  _{i+j}-\mu _{i}\mu _{j}\right) .
\end{eqnarray}

The {\em absolute} error is then again given by
$\Delta C_{n} = \sqrt{\var\left( C_{n} \right) }$.   
For example, we obtain the following for the variance of $C_{2}$ (after re-expressing the moments $\mu_{i}$ in terms of
factorial cumulants)
\begin{align}
  \var\left( C_{2} \right) = \frac{1}{\neve} &\left[ 2 \left(C_1+C_2\right)^{2} + 2 C_2 
  + 4C_3 + C_4 \right].
  \label{eq:varC2formula}
\end{align}
We find that the so obtained errors are in perfect agreement with those determined via the aforementioned
numerical sampling method.

\begin{figure}[t]
\begin{center}
  \includegraphics[scale=0.35]{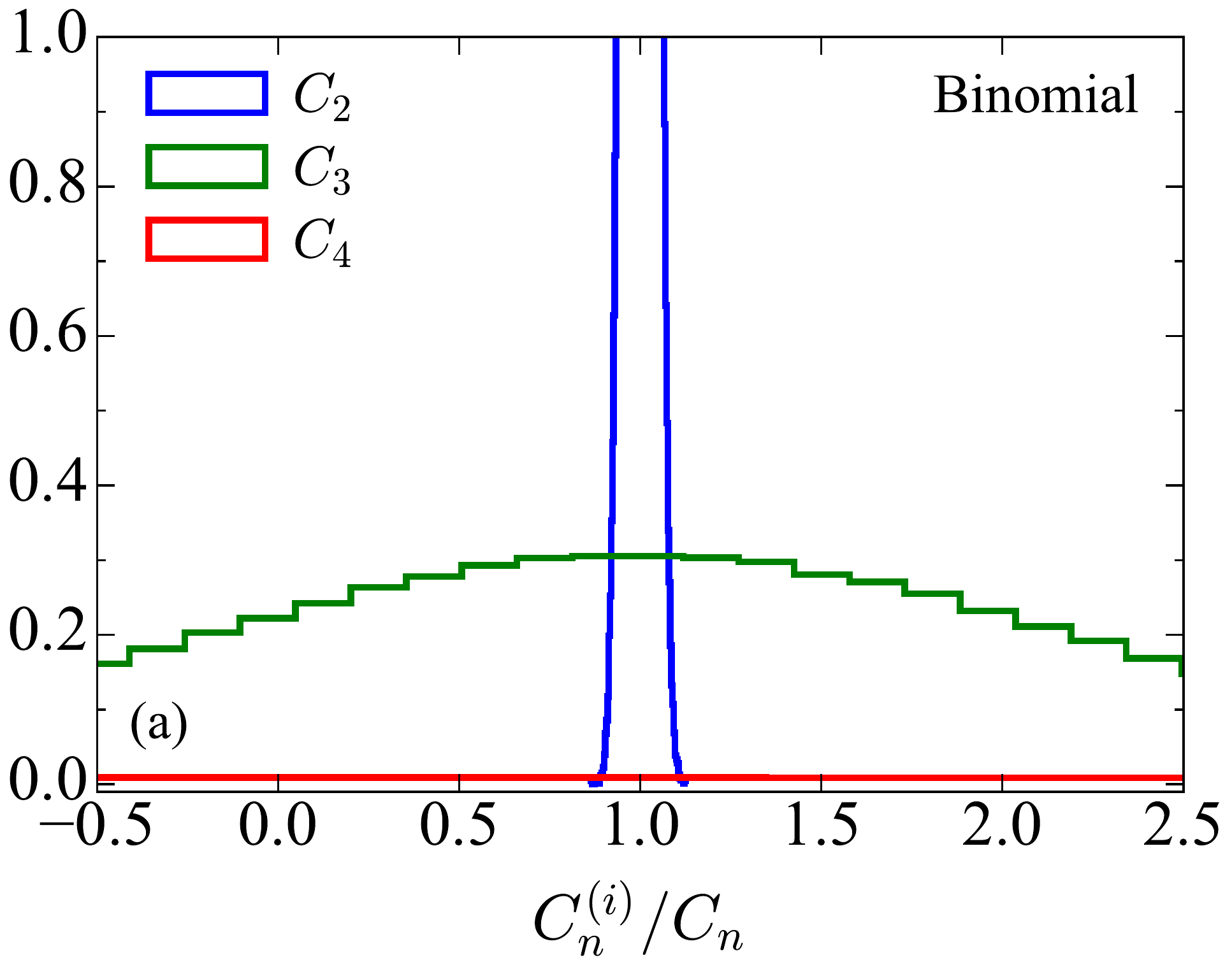}    

  \vspace{3mm}
  
  \includegraphics[scale=0.35]{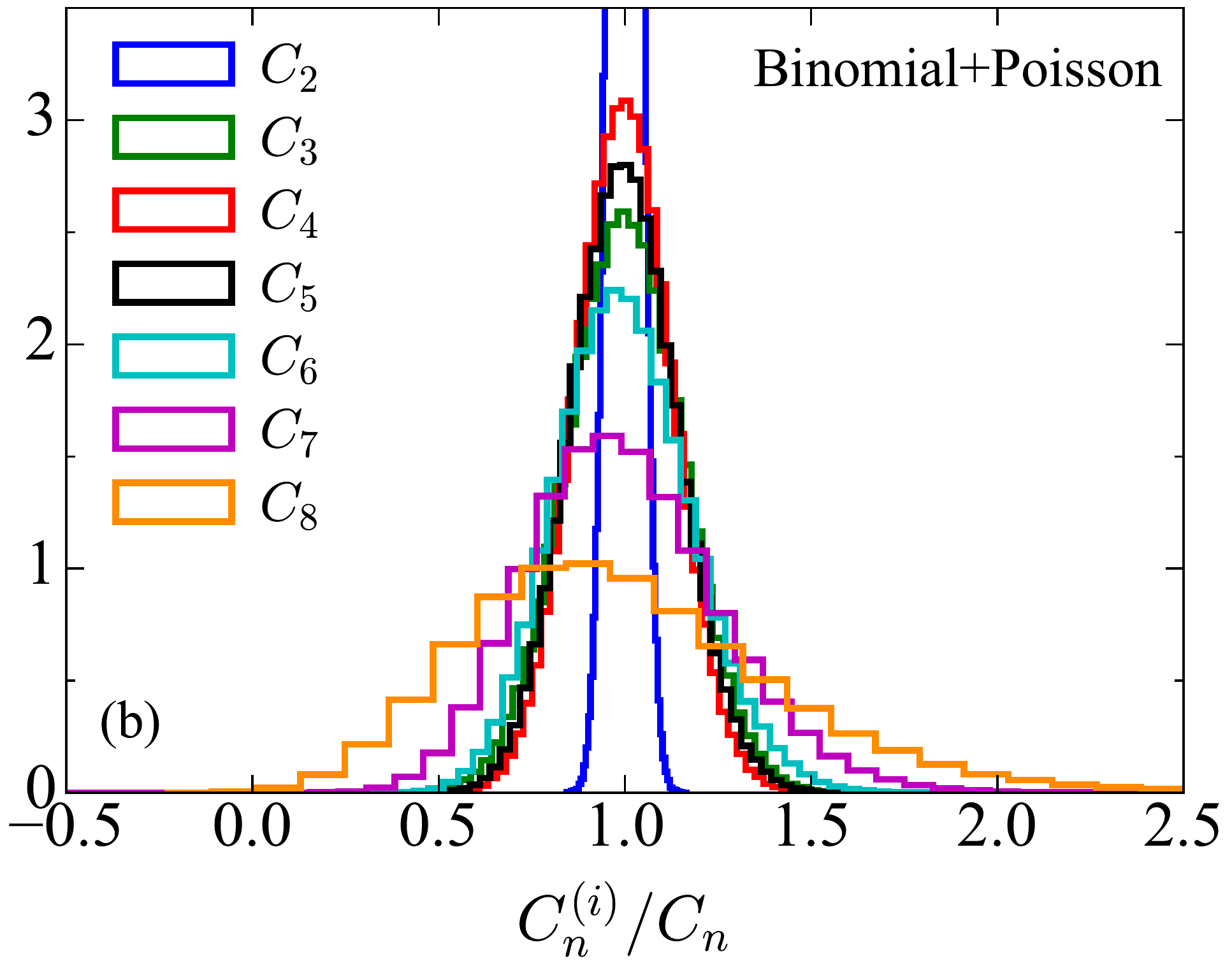}
\end{center}
\par
\vspace{-5mm}
\caption{Histogram (normalized to unity) of the factorial cumulant, $C_{n}^{(i)}$, fluctuating from experiment 
  to experiment, divided by a known (evaluated analytically) value, $C_{n}$, based on 144393 events sampled from (a) the
  binomial distribution ($B=350$, $p=0.114$, $\langle N \rangle = pB \approx 40$) and (b) a distribution given 
  by Eq. (\ref{eq:two_component}) (see text for details). The statistical
  errors are given by the widths of the corresponding
  histograms. In the panel (a), the histograms' order by the height at the maximum is (from largest
  to smallest) $C_{2}$,
  $C_{3}$, and $C_{4}$. In the panel (b) the order (from largest to smallest) is: $C_{2}$,
  $C_{4}$, $C_{5}$, $C_{3}$, $C_{6}$, $C_{7}$, and $C_{8}$.}
\label{fig:Bino}
\end{figure}

After having presented the methods for error determination let us turn to the results. The
essential point of the present paper is the observation that a small deviation from Poisson or
binomial distributions can result in rather peculiar distributions, which we call
statistics friendly distributions. From these distributions one may obtain factorial cumulants of high orders with a rather limited number of events. One example is a simple two-component distribution discussed recently in Ref. \cite{Bzdak:2018uhv}
\begin{equation}
P(N)=(1-\alpha )P_{(a)}(N)+\alpha P_{(b)}(N),
\label{eq:two_component}
\end{equation}
where both $P_{(a)}$  
and $P_{(b)}$ are the proton multiplicity distributions characterized by small or even vanishing
factorial cumulants.\footnote{The simplest two-component distribution could result from two Poissons 
with different means. We take the main distribution to be binomial to conserve the baryon number, however, 
this is not important for our conclusions.} This distribution not only serves 
as a nice example for a statistics friendly distribution, but also, 
as argued recently in Ref.~\cite{Bzdak:2018uhv}, such a distribution would be consistent with a
system with a finite number of particles being close to a first-order phase transition.
The analysis in Ref.~\cite{Bzdak:2018uhv} found that $P(N)$ given by
Eq. (\ref{eq:two_component}) with $\alpha \approx 0.0033$, $P_{(a)}(N)$ given by binomial ($B=350$,
$p\approx0.1144$) and $P_{(b)}(N)$ given by Poisson ($\langle N_{(b)} \rangle = 25.3525$) is able
to reproduce the preliminary results by the STAR Collaboration for the proton cumulants  at
$\sqrt{s}=7.7\gev$ \cite{Luo:2015ewa}. In addition it was found that the above distribution predicts
factorial cumulants to roughly scale like $C_{n+1}/C_{n} \sim -15$, i.e., they alter in sign from order to order
while increasing in absolute value by more than an order of magnitude.\footnote{The actual ratios
  slightly decrease with increasing $n$: $C_{4}/C_{3}=-17$, $C_{5}/C_{4}=-15.56$, 
  $C_{6}/C_{5}=-15.46$, ${C_{7}/C_{6}=-15.04}$, $C_{8}/C_{7}=-13.85$, $C_{9}/C_{8}=-10.66$, 
  and for $C_{10}$ the pattern breaks and $C_{10}/C_{9}=0.72$.} In other words, the small
admixture of a Poisson distribution changes the factorial cumulants dramatically, from being close
to zero to almost exponentially increasing in magnitude. The same dramatic difference  can also be seen in
the expected error for a finite sampling. This is shown in Fig. \ref{fig:Bino}, where in
panel (a) we show the histogram of $C_{n}^{(i)}/C_{n}$ from our numerical sampling (based
on $\neve=\neveN$ events) for the binomial distribution only, 
i.e., $P_{(a)}(N)$.\footnote{We found that the absolute error of $C_{n}$ for the
  binomial distribution is close to that of the Poisson distribution, which can be easily calculated using Eq. (\ref{eq:delta}) and is given by 
\unexpanded{$\sqrt{n!}\langle N \rangle^{n/2}/\sqrt{\neve}$.}} For completeness we note that the analytical values of $C_{n}$ for binomial are given by $C_{n}=(-1)^{n+1}(n-1)! B p^{n}$. The distribution gets very wide already for $C_3$. 
In contrast, in  panel (b) of Fig. \ref{fig:Bino} we show equivalent
histograms for the two-component distribution, Eq.~\eqref{eq:two_component}. Again, the small
admixture of a Poisson distribution changes the situation dramatically. In this case the
distributions are so narrow that a measurement of even the 8-th order factorial cumulants may be
feasible with as little as 150000 events. 

\begin{figure}[t]
\begin{center}
\includegraphics[scale=0.35]{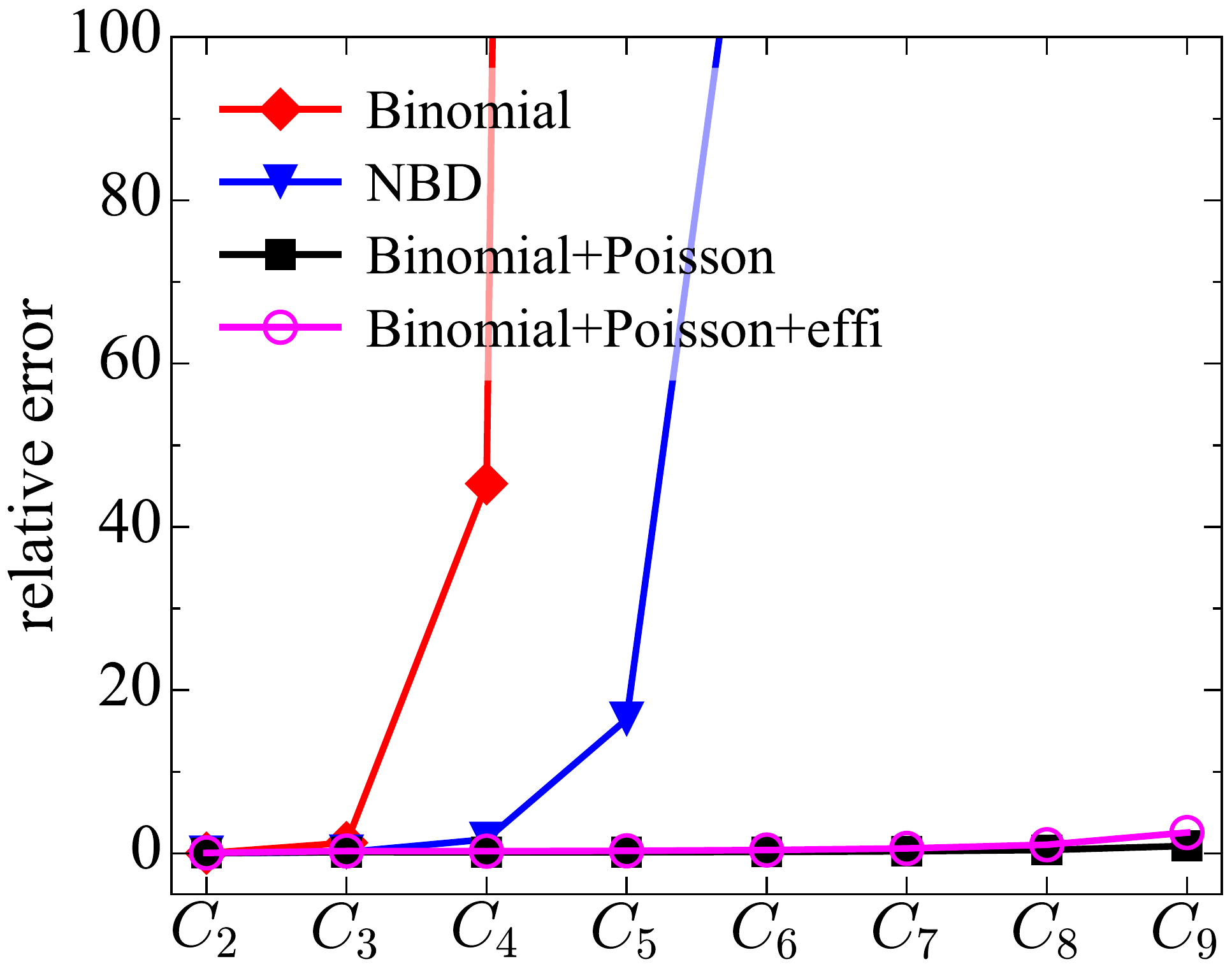}
\end{center}
\par
\vspace{-5mm}
\caption{The relative error, $\Delta C_{n}/C_{n}$, of factorial cumulants for various proton multiplicity distributions based on 144393 events, as present in the most central $Au+Au$ collisions at RHIC. The binomial and negative binomial distributions presented here are statistically very demanding, whereas the distribution given by. Eq. (\ref{eq:two_component}) (Binomial+Poisson) with $\avet{N}=40$, allows to successfully measure higher order factorial cumulants with a relatively small number of events. This feature is also present for the efficiency uncorrected distribution (Binomial+Poisson+effi) where $\avet{N}=40\times0.65$.}
\label{fig:width}
\end{figure}

This finding is quantified in Fig.~\ref{fig:width}, where we show the {\em relative} errors
$\Delta C_{n}/C_{n}$ for 
various distributions again based on $\neve=\neveN$ event. 
The relative error for both the binomial distribution and the negative binomial distribution (NBD)\footnote{For the NBD \unexpanded{$C_{n}=(n-1)! \langle N \rangle^{n}/k^{n-1}$}, where $k$ measures the deviation from a Poisson distribution, e.g., 
\unexpanded{$\langle N^{2} \rangle - \langle N \rangle^2 = \langle N \rangle (1 + \langle N \rangle /k)$}.} with
$\langle N \rangle = 40$ and $k=80$, increase essentially exponentially with increasing order of the
factorial cumulant.
Obviously all of these distributions are statistics hungry, and the measurement of higher order factorial cumulants with  good accuracy requires very large statistics.
For the  two-component model, labeled ``Binomial + Poisson'', on the other hand
the relative errors remain very small even for $C_{9}$. The actual values for the relative errors are
(0.036, 0.16, 0.13, 0.14, 0.18, 0.26, 0.42, 0.91) for $(\Delta C_{2}/C_{2}, \ldots, \Delta C_{9}/C_{9})$.

We also show as  ``Binomial + Poisson +
effi'' the result one would obtain, if one takes a finite detection efficiency of $\epsilon =0.65$
into account, 
that is $\ave{N_{(b)}}=25.3525\times\epsilon$, $p=0.1144\times \epsilon$ so that $
\ave{N}=40\times\epsilon$. Again, the relative error for the factorial cumulants remains small but
larger than that in the case without efficiency. Here we have  (0.056, 0.29, 0.27, 0.31,
0.41, 0.61, 1.06, 2.55) for $(\Delta C_{2}/C_{2}, \ldots, \Delta C_{9}/C_{9})$. This also
means that using the efficiency uncorrected STAR data one could try to measure the factorial
cumulants up to the seventh order where $\Delta C_7/C_{7}=1\pm0.61$.

The above results may be understood qualitatively in the following way. In general we have two types
of multiplicity distributions, $P(N)$.  
One where the higher order factorial cumulants are driven by the tails (Poisson, binomial, NBD etc.)
and the other one where the higher order factorial cumulants are driven by some structure away from
the tails. This is exactly the case of our model.\footnote{Another example of a statistics friendly distribution is a uniform distribution. For example, taking $P(N)=$ const for $N\in[0,80]$ we obtain 
(0.0026, 0.0309, 0.0071, 0.0447, 0.0114, 0.0477, 0.0157, 0.0487) 
for $(\Delta C_{2}/C_{2}, \ldots, \Delta C_{9}/C_{9})$.}
To be a bit more precise the factorial cumulants of Eq. (\ref{eq:two_component}), assuming $\alpha \ll 1$ are given by\footnote{Again, we assume that both $P_{(a)}$  
and $P_{(b)}$ are the proton multiplicity distributions characterized by small (or even vanishing)
factorial cumulants. The whole idea is to obtain large factorial cumulants from two rather standard distributions.}
\begin{equation}
C_{n}\, \approx \, C_{n}^{(a)} + (-1)^{n}\alpha\nb^{n} ,
\label{eq:Cn_two_component}
\end{equation}
where $C_{n}^{(a)}$ is a factorial cumulant characterizing $P_{(a)}(N)$ and 
$\nb = \langle N_{(a)}\rangle - \langle N_{(b)}\rangle$. For $C_{n}^{(a)}$ being a Poisson or binomial
the values of $C_{n}$ are completely dominated by the term $\alpha\nb^{n}$, which results in very
large factorial cumulants. The error, $\Delta C_{n}$, on the other hand, is of the same magnitude as
that of the first term, $\Delta C^{(a)}_{n}$ (in practice
$\Delta C_{n}^{(a)} / \Delta C_{n}$ ranges from  $\sim 0.95$ for $n=2$ to $\sim 0.2$ for $n=9$). 
Thus we have a situation, where the error of 
the factorial cumulant is of the same
magnitude as that of a binomial distribution, but the factorial cumulant is orders of magnitude
larger. Consequently, and not surprisingly, the {\em relative} error is much smaller for
the two-component distribution than for the binomial distribution. It is worth noting that $C_n$ 
scales linearly with $\alpha$ and the two-component distribution is statistics friendly even if the 
second mode is tiny, i.e., $\alpha$ is small (provided $\nb^{n}$ is large enough).

\begin{figure}[t]
\begin{center}
\includegraphics[scale=0.35]{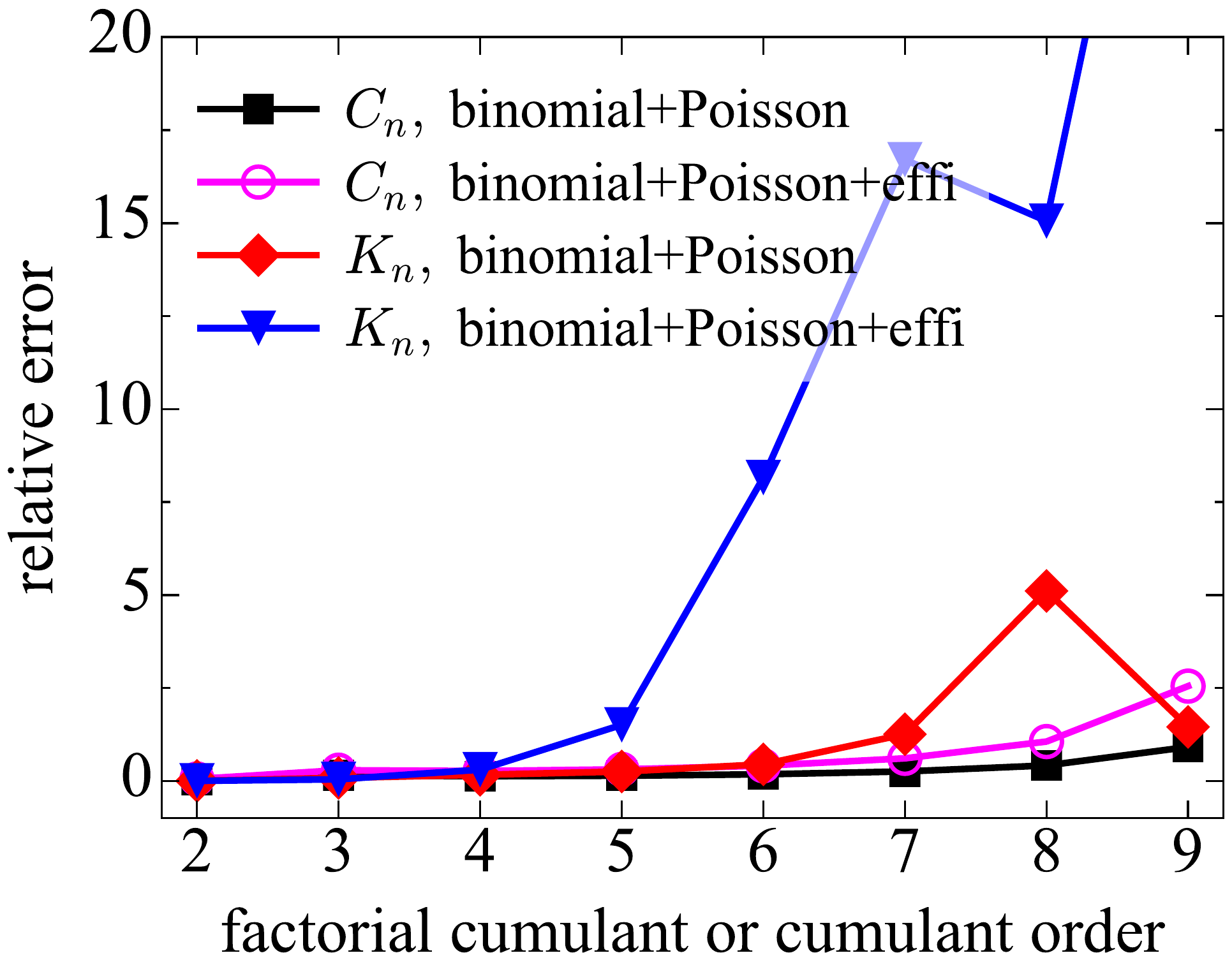}
\end{center}
\par
\vspace{-5mm}
\caption{The relative errors of the factorial cumulants, $\Delta C_{n}/C_{n}$, 
and the regular cumulants, $\Delta K_{n}/K_{n}$, based on 144393 events sampled from 
a distribution given by Eq. (\ref{eq:two_component}).}
\label{fig:width_comparison}
\end{figure}

Finally, we note that in the case of Eq. (\ref{eq:two_component}), the regular cumulnats are less
statistics friendly. This is presented in Fig. \ref{fig:width_comparison}. The reason for this is
the same as just stated. The absolute errors for both cumulants and factorial cumulants are of
the same magnitude, $\Delta K_{n} \sim \Delta C_{n}$. On the other hand, for the two-component
model, the factorial cumulants are very large
while the regular cumulants are only modestly larger than that of a simple binomial distribution. This is
a result of the alternating signs of the factorial cumulants. For example, the
sixth order cumulant, $K_{6}$, is given in terms of the factorial cumulants as  $K_{6}= \langle
N\rangle + 31 C_{2} + 90 C_{3} + 65 C_{4} + 15 C_{5} + C_{6}$ (see e.g.,
Ref. \cite{Bzdak:2018uhv}). For our example of "binomial+Poisson+effi", where we see a rapid
increase in the relative error, we have $C_{6} \approx 3080$, $15 C_{5} \approx -4600$ 
and $65 C_{4} \approx 1970$. As a result, $K_{6}\approx 180\ll C_{6}$, and consequently 
the relative error is much larger for $K_{6}$ as compared to $C_{6}$. 

In summary, we demonstrated that for the multiplicity distribution given by
Eq. (\ref{eq:two_component}), which is relevant in the context of
searching for structures in the QCD phase diagram, factorial cumulants of high orders can be 
determined with a relatively small
number of events.  This is in contrast to various statistics hungry distributions (Poisson,
binomial, NBD, etc.), for which the error increases nearly exponentially with increasing order. 
As shown in Ref.~\cite{Bzdak:2018uhv}, the  distribution, Eq.~\eqref{eq:two_component}, describes
the preliminary STAR data for proton cumulants (up to the forth order) in central Au+Au collisions
at $\sqrt{s}=7.7\gev$. Because this distribution is statistics friendly, it can be further tested
by evaluating the higher order factorial cumulants even with the presently available STAR data set
of $\neveN$ events for the most central collisions.   
We also pointed out that factorial cumulants are more statistics friendly when compared to regular
cumulants, which, in the case of Eq. (\ref{eq:two_component}), results from a delicate cancellation
of large factorial cumulants. Assuming that $C_4=170$ (as extracted from preliminary STAR data) we
predict:
\begin{eqnarray}
C_{5} &=& -307 \, (1\pm 0.31), \quad    C_{6} = 3085 \, (1\pm 0.41), \notag \\
C_{7} &=& -30155 \, (1\pm 0.61), \quad  C_{8} = 271492 \, (1\pm 1.06), \notag
\end{eqnarray}%
for efficiency uncorrected data and 
\begin{eqnarray}
C_{5} &=& -2645 \, (1\pm 0.14), \quad    C_{6} = 40900 \, (1\pm 0.18), \notag \\
C_{7} &=& -615135 \, (1\pm 0.26), \quad  C_{8} =8520220 \, (1\pm 0.42), \notag
\end{eqnarray}%
for $\langle N\rangle = 40$, corresponding to the efficiency corrected data.\footnote{We note that
  the errors quoted here are only due to the sample size and do not account for additional
  uncertainties due to the efficiency correction \cite{Luo:2014rea}.} In the next
phase of the RHIC beam energy scan the statistics is expected to increase by roughly a factor of
$\sim 25$ \cite{star_note} reducing the above errors by about a factor of 5.
It would be desirable to also analyze $C_{5}$ and higher order proton (not net-proton) 
  factorial cumulants at much higher energies, say,
  $\sqrt{s} = 200$ GeV, where a first-order phase transition is not anticipated. Thus
  the factorial cumulants are not expected to alter in sign while increasing in absolute
  value. It was checked in Ref. \cite{Bzdak:2016sxg} that $C_3$ and $C_4$ alter in sign but their 
  magnitudes are very small.

Our message does not rely on the ability to  estimate the errors of $C_{n}$ in an
experiment. The reason is the following. We conjecture that the multiplicity distribution at 7.7 GeV
is a two-component one and describe the preliminary data up to the fourth order. Next, we run a sufficient
number of independent experiments with each experiment resulting in one measured number $C_{n}$. The
histogram of the measured values, as shown in Fig.~\ref{fig:Bino}(b), is narrow if the distribution is
given by our conjectured one. Now STAR makes one measurement only and obtains $C_5$, $C_6$, $C_7$,
$C_8$. If our conjecture is correct, that is, the distribution is a two-component one, the numbers
measured by STAR should be consistent with our predictions. If the numbers are significantly off our
predictions, then our conjecture is falsified. We also note that this procedure is quite general and
not restricted to the STAR data discussed here: Measure the first four factorial cumulants then 
see if they are consistent with a two-component distribution. If so, test this distribution by
comparing the measured higher factorial cumulants with the prediction of the two-component model.

In conclusion, we have shown that two-component multiplicity distributions as expected in the
vicinity of a first-order phase transition are ``statistics-friendly''. This allows for the
determination of factorial cumulants of high order even with limited statistics, and opens a novel
way to search for the phase structure of mesoscopic systems.

\bigskip
{\bf Acknowledgments:} We thank Andrzej Bialas and Jan Steinheimer for useful comments. 
A.B. is partially supported by the Ministry of Science and Higher Education, and by the National 
Science Centre Grant No. 2018/30/Q/ST2/00101. 
V.K. is supported by the U.S. Department of Energy, Office of Science, Office of Nuclear Physics, 
under contract number DE-AC02-05CH11231. 
This work also received support within the framework of the Beam Energy Scan Theory (BEST) 
Topical Collaboration.

\bibliography{paper}

\begin{thebibliography}{31}%
\makeatletter
\providecommand \@ifxundefined [1]{%
 \@ifx{#1\undefined}
}%
\providecommand \@ifnum [1]{%
 \ifnum #1\expandafter \@firstoftwo
 \else \expandafter \@secondoftwo
 \fi
}%
\providecommand \@ifx [1]{%
 \ifx #1\expandafter \@firstoftwo
 \else \expandafter \@secondoftwo
 \fi
}%
\providecommand \natexlab [1]{#1}%
\providecommand \enquote  [1]{``#1''}%
\providecommand \bibnamefont  [1]{#1}%
\providecommand \bibfnamefont [1]{#1}%
\providecommand \citenamefont [1]{#1}%
\providecommand \href@noop [0]{\@secondoftwo}%
\providecommand \href [0]{\begingroup \@sanitize@url \@href}%
\providecommand \@href[1]{\@@startlink{#1}\@@href}%
\providecommand \@@href[1]{\endgroup#1\@@endlink}%
\providecommand \@sanitize@url [0]{\catcode `\\12\catcode `\$12\catcode
  `\&12\catcode `\#12\catcode `\^12\catcode `\_12\catcode `\%12\relax}%
\providecommand \@@startlink[1]{}%
\providecommand \@@endlink[0]{}%
\providecommand \url  [0]{\begingroup\@sanitize@url \@url }%
\providecommand \@url [1]{\endgroup\@href {#1}{\urlprefix }}%
\providecommand \urlprefix  [0]{URL }%
\providecommand \Eprint [0]{\href }%
\providecommand \doibase [0]{http://dx.doi.org/}%
\providecommand \selectlanguage [0]{\@gobble}%
\providecommand \bibinfo  [0]{\@secondoftwo}%
\providecommand \bibfield  [0]{\@secondoftwo}%
\providecommand \translation [1]{[#1]}%
\providecommand \BibitemOpen [0]{}%
\providecommand \bibitemStop [0]{}%
\providecommand \bibitemNoStop [0]{.\EOS\space}%
\providecommand \EOS [0]{\spacefactor3000\relax}%
\providecommand \BibitemShut  [1]{\csname bibitem#1\endcsname}%
\let\auto@bib@innerbib\@empty
\bibitem [{\citenamefont {Aoki}\ \emph {et~al.}(2006)\citenamefont {Aoki},
  \citenamefont {Endrodi}, \citenamefont {Fodor}, \citenamefont {Katz},\ and\
  \citenamefont {Szabo}}]{Aoki:2006we}%
  \BibitemOpen
  \bibfield  {author} {\bibinfo {author} {\bibfnamefont {Y.}~\bibnamefont
  {Aoki}}, \bibinfo {author} {\bibfnamefont {G.}~\bibnamefont {Endrodi}},
  \bibinfo {author} {\bibfnamefont {Z.}~\bibnamefont {Fodor}}, \bibinfo
  {author} {\bibfnamefont {S.~D.}\ \bibnamefont {Katz}}, \ and\ \bibinfo
  {author} {\bibfnamefont {K.~K.}\ \bibnamefont {Szabo}},\ }\href {\doibase
  10.1038/nature05120} {\bibfield  {journal} {\bibinfo  {journal} {Nature}\
  }\textbf {\bibinfo {volume} {443}},\ \bibinfo {pages} {675} (\bibinfo {year}
  {2006})},\ \Eprint {http://arxiv.org/abs/hep-lat/0611014}
  {arXiv:hep-lat/0611014} \BibitemShut {NoStop}%
\bibitem [{\citenamefont {Stephanov}(2004)}]{Stephanov:2004wx}%
  \BibitemOpen
  \bibfield  {author} {\bibinfo {author} {\bibfnamefont {M.~A.}\ \bibnamefont
  {Stephanov}},\ }\href@noop {} {\bibfield  {journal} {\bibinfo  {journal}
  {Prog. Theor. Phys. Suppl.}\ }\textbf {\bibinfo {volume} {153}},\ \bibinfo
  {pages} {139} (\bibinfo {year} {2004})},\ \Eprint
  {http://arxiv.org/abs/hep-ph/0402115} {arXiv:hep-ph/0402115} \BibitemShut
  {NoStop}%
\bibitem [{\citenamefont {Bzdak}\ \emph {et~al.}(2019)\citenamefont {Bzdak},
  \citenamefont {Esumi}, \citenamefont {Koch}, \citenamefont {Liao},
  \citenamefont {Stephanov},\ and\ \citenamefont {Xu}}]{Bzdak:2019pkr}%
  \BibitemOpen
  \bibfield  {author} {\bibinfo {author} {\bibfnamefont {A.}~\bibnamefont
  {Bzdak}}, \bibinfo {author} {\bibfnamefont {S.}~\bibnamefont {Esumi}},
  \bibinfo {author} {\bibfnamefont {V.}~\bibnamefont {Koch}}, \bibinfo {author}
  {\bibfnamefont {J.}~\bibnamefont {Liao}}, \bibinfo {author} {\bibfnamefont
  {M.}~\bibnamefont {Stephanov}}, \ and\ \bibinfo {author} {\bibfnamefont
  {N.}~\bibnamefont {Xu}},\ }\href@noop {} {\  (\bibinfo {year} {2019})},\
  \Eprint {http://arxiv.org/abs/1906.00936} {arXiv:1906.00936 [nucl-th]}
  \BibitemShut {NoStop}%
\bibitem [{\citenamefont {Jeon}\ and\ \citenamefont
  {Koch}(2000)}]{Jeon:2000wg}%
  \BibitemOpen
  \bibfield  {author} {\bibinfo {author} {\bibfnamefont {S.}~\bibnamefont
  {Jeon}}\ and\ \bibinfo {author} {\bibfnamefont {V.}~\bibnamefont {Koch}},\
  }\href {\doibase 10.1103/PhysRevLett.85.2076} {\bibfield  {journal} {\bibinfo
   {journal} {Phys. Rev. Lett.}\ }\textbf {\bibinfo {volume} {85}},\ \bibinfo
  {pages} {2076} (\bibinfo {year} {2000})},\ \Eprint
  {http://arxiv.org/abs/hep-ph/0003168} {arXiv:hep-ph/0003168 [hep-ph]}
  \BibitemShut {NoStop}%
\bibitem [{\citenamefont {Asakawa}\ \emph {et~al.}(2000)\citenamefont
  {Asakawa}, \citenamefont {Heinz},\ and\ \citenamefont
  {Muller}}]{Asakawa:2000wh}%
  \BibitemOpen
  \bibfield  {author} {\bibinfo {author} {\bibfnamefont {M.}~\bibnamefont
  {Asakawa}}, \bibinfo {author} {\bibfnamefont {U.~W.}\ \bibnamefont {Heinz}},
  \ and\ \bibinfo {author} {\bibfnamefont {B.}~\bibnamefont {Muller}},\
  }\href@noop {} {\bibfield  {journal} {\bibinfo  {journal} {Phys. Rev. Lett.}\
  }\textbf {\bibinfo {volume} {85}},\ \bibinfo {pages} {2072} (\bibinfo {year}
  {2000})},\ \Eprint {http://arXiv.org/abs/hep-ph/0003169} {hep-ph/0003169}
  \BibitemShut {NoStop}%
\bibitem [{\citenamefont {Stephanov}(2009)}]{Stephanov:2008qz}%
  \BibitemOpen
  \bibfield  {author} {\bibinfo {author} {\bibfnamefont {M.}~\bibnamefont
  {Stephanov}},\ }\href {\doibase 10.1103/PhysRevLett.102.032301} {\bibfield
  {journal} {\bibinfo  {journal} {Phys. Rev. Lett.}\ }\textbf {\bibinfo
  {volume} {102}},\ \bibinfo {pages} {032301} (\bibinfo {year} {2009})},\
  \Eprint {http://arxiv.org/abs/0809.3450} {arXiv:0809.3450 [hep-ph]}
  \BibitemShut {NoStop}%
\bibitem [{\citenamefont {Skokov}\ \emph {et~al.}(2011)\citenamefont {Skokov},
  \citenamefont {Friman},\ and\ \citenamefont {Redlich}}]{Skokov:2010uh}%
  \BibitemOpen
  \bibfield  {author} {\bibinfo {author} {\bibfnamefont {V.}~\bibnamefont
  {Skokov}}, \bibinfo {author} {\bibfnamefont {B.}~\bibnamefont {Friman}}, \
  and\ \bibinfo {author} {\bibfnamefont {K.}~\bibnamefont {Redlich}},\ }\href
  {\doibase 10.1103/PhysRevC.83.054904} {\bibfield  {journal} {\bibinfo
  {journal} {Phys. Rev.}\ }\textbf {\bibinfo {volume} {C83}},\ \bibinfo {pages}
  {054904} (\bibinfo {year} {2011})},\ \Eprint {http://arxiv.org/abs/1008.4570}
  {arXiv:1008.4570 [hep-ph]} \BibitemShut {NoStop}%
\bibitem [{\citenamefont {Stephanov}(2011)}]{Stephanov:2011pb}%
  \BibitemOpen
  \bibfield  {author} {\bibinfo {author} {\bibfnamefont {M.}~\bibnamefont
  {Stephanov}},\ }\href {\doibase 10.1103/PhysRevLett.107.052301} {\bibfield
  {journal} {\bibinfo  {journal} {Phys. Rev. Lett.}\ }\textbf {\bibinfo
  {volume} {107}},\ \bibinfo {pages} {052301} (\bibinfo {year} {2011})},\
  \Eprint {http://arxiv.org/abs/1104.1627} {arXiv:1104.1627 [hep-ph]}
  \BibitemShut {NoStop}%
\bibitem [{\citenamefont {Luo}\ \emph {et~al.}(2012)\citenamefont {Luo},
  \citenamefont {Mohanty}, \citenamefont {Ritter},\ and\ \citenamefont
  {Xu}}]{Luo:2011rg}%
  \BibitemOpen
  \bibfield  {author} {\bibinfo {author} {\bibfnamefont {X.-F.}\ \bibnamefont
  {Luo}}, \bibinfo {author} {\bibfnamefont {B.}~\bibnamefont {Mohanty}},
  \bibinfo {author} {\bibfnamefont {H.~G.}\ \bibnamefont {Ritter}}, \ and\
  \bibinfo {author} {\bibfnamefont {N.}~\bibnamefont {Xu}},\ }\href {\doibase
  10.1134/S1063778812060348} {\bibfield  {journal} {\bibinfo  {journal} {Phys.
  Atom. Nucl.}\ }\textbf {\bibinfo {volume} {75}},\ \bibinfo {pages} {676}
  (\bibinfo {year} {2012})},\ \Eprint {http://arxiv.org/abs/1105.5049}
  {arXiv:1105.5049 [nucl-ex]} \BibitemShut {NoStop}%
\bibitem [{\citenamefont {Luo}\ and\ \citenamefont {Xu}(2017)}]{Luo:2017faz}%
  \BibitemOpen
  \bibfield  {author} {\bibinfo {author} {\bibfnamefont {X.}~\bibnamefont
  {Luo}}\ and\ \bibinfo {author} {\bibfnamefont {N.}~\bibnamefont {Xu}},\
  }\href {\doibase 10.1007/s41365-017-0257-0} {\bibfield  {journal} {\bibinfo
  {journal} {Nucl. Sci. Tech.}\ }\textbf {\bibinfo {volume} {28}},\ \bibinfo
  {pages} {112} (\bibinfo {year} {2017})},\ \Eprint
  {http://arxiv.org/abs/1701.02105} {arXiv:1701.02105 [nucl-ex]} \BibitemShut
  {NoStop}%
\bibitem [{\citenamefont {Herold}\ \emph {et~al.}(2016)\citenamefont {Herold},
  \citenamefont {Nahrgang}, \citenamefont {Yan},\ and\ \citenamefont
  {Kobdaj}}]{Herold:2016uvv}%
  \BibitemOpen
  \bibfield  {author} {\bibinfo {author} {\bibfnamefont {C.}~\bibnamefont
  {Herold}}, \bibinfo {author} {\bibfnamefont {M.}~\bibnamefont {Nahrgang}},
  \bibinfo {author} {\bibfnamefont {Y.}~\bibnamefont {Yan}}, \ and\ \bibinfo
  {author} {\bibfnamefont {C.}~\bibnamefont {Kobdaj}},\ }\href {\doibase
  10.1103/PhysRevC.93.021902} {\bibfield  {journal} {\bibinfo  {journal} {Phys.
  Rev.}\ }\textbf {\bibinfo {volume} {C93}},\ \bibinfo {pages} {021902}
  (\bibinfo {year} {2016})},\ \Eprint {http://arxiv.org/abs/1601.04839}
  {arXiv:1601.04839 [hep-ph]} \BibitemShut {NoStop}%
\bibitem [{\citenamefont {Zhou}\ \emph {et~al.}(2012)\citenamefont {Zhou},
  \citenamefont {Limphirat}, \citenamefont {Yan}, \citenamefont {Yun},
  \citenamefont {Yan}, \citenamefont {Cai}, \citenamefont {Csernai},\ and\
  \citenamefont {Sa}}]{Zhou:2012ay}%
  \BibitemOpen
  \bibfield  {author} {\bibinfo {author} {\bibfnamefont {D.-M.}\ \bibnamefont
  {Zhou}}, \bibinfo {author} {\bibfnamefont {A.}~\bibnamefont {Limphirat}},
  \bibinfo {author} {\bibfnamefont {Y.-l.}\ \bibnamefont {Yan}}, \bibinfo
  {author} {\bibfnamefont {C.}~\bibnamefont {Yun}}, \bibinfo {author}
  {\bibfnamefont {Y.-p.}\ \bibnamefont {Yan}}, \bibinfo {author} {\bibfnamefont
  {X.}~\bibnamefont {Cai}}, \bibinfo {author} {\bibfnamefont {L.~P.}\
  \bibnamefont {Csernai}}, \ and\ \bibinfo {author} {\bibfnamefont {B.-H.}\
  \bibnamefont {Sa}},\ }\href {\doibase 10.1103/PhysRevC.85.064916} {\bibfield
  {journal} {\bibinfo  {journal} {Phys. Rev.}\ }\textbf {\bibinfo {volume}
  {C85}},\ \bibinfo {pages} {064916} (\bibinfo {year} {2012})},\ \Eprint
  {http://arxiv.org/abs/1205.5634} {arXiv:1205.5634 [nucl-th]} \BibitemShut
  {NoStop}%
\bibitem [{\citenamefont {Wang}\ and\ \citenamefont
  {Yang}(2012)}]{Wang:2012jr}%
  \BibitemOpen
  \bibfield  {author} {\bibinfo {author} {\bibfnamefont {X.}~\bibnamefont
  {Wang}}\ and\ \bibinfo {author} {\bibfnamefont {C.~B.}\ \bibnamefont
  {Yang}},\ }\href {\doibase 10.1103/PhysRevC.85.044905} {\bibfield  {journal}
  {\bibinfo  {journal} {Phys. Rev.}\ }\textbf {\bibinfo {volume} {C85}},\
  \bibinfo {pages} {044905} (\bibinfo {year} {2012})},\ \Eprint
  {http://arxiv.org/abs/1202.4857} {arXiv:1202.4857 [nucl-th]} \BibitemShut
  {NoStop}%
\bibitem [{\citenamefont {Karsch}\ and\ \citenamefont
  {Redlich}(2011)}]{Karsch:2011gg}%
  \BibitemOpen
  \bibfield  {author} {\bibinfo {author} {\bibfnamefont {F.}~\bibnamefont
  {Karsch}}\ and\ \bibinfo {author} {\bibfnamefont {K.}~\bibnamefont
  {Redlich}},\ }\href {\doibase 10.1103/PhysRevD.84.051504} {\bibfield
  {journal} {\bibinfo  {journal} {Phys. Rev.}\ }\textbf {\bibinfo {volume}
  {D84}},\ \bibinfo {pages} {051504} (\bibinfo {year} {2011})},\ \Eprint
  {http://arxiv.org/abs/1107.1412} {arXiv:1107.1412 [hep-ph]} \BibitemShut
  {NoStop}%
\bibitem [{\citenamefont {Schaefer}\ and\ \citenamefont
  {Wagner}(2012)}]{Schaefer:2011ex}%
  \BibitemOpen
  \bibfield  {author} {\bibinfo {author} {\bibfnamefont {B.~J.}\ \bibnamefont
  {Schaefer}}\ and\ \bibinfo {author} {\bibfnamefont {M.}~\bibnamefont
  {Wagner}},\ }\href {\doibase 10.1103/PhysRevD.85.034027} {\bibfield
  {journal} {\bibinfo  {journal} {Phys. Rev.}\ }\textbf {\bibinfo {volume}
  {D85}},\ \bibinfo {pages} {034027} (\bibinfo {year} {2012})},\ \Eprint
  {http://arxiv.org/abs/1111.6871} {arXiv:1111.6871 [hep-ph]} \BibitemShut
  {NoStop}%
\bibitem [{\citenamefont {Chen}\ \emph {et~al.}(2011)\citenamefont {Chen},
  \citenamefont {Pan}, \citenamefont {Xiong}, \citenamefont {Li}, \citenamefont
  {Li}, \citenamefont {Li}, \citenamefont {Wang},\ and\ \citenamefont
  {Wu}}]{Chen:2011am}%
  \BibitemOpen
  \bibfield  {author} {\bibinfo {author} {\bibfnamefont {L.}~\bibnamefont
  {Chen}}, \bibinfo {author} {\bibfnamefont {X.}~\bibnamefont {Pan}}, \bibinfo
  {author} {\bibfnamefont {F.-B.}\ \bibnamefont {Xiong}}, \bibinfo {author}
  {\bibfnamefont {L.}~\bibnamefont {Li}}, \bibinfo {author} {\bibfnamefont
  {N.}~\bibnamefont {Li}}, \bibinfo {author} {\bibfnamefont {Z.}~\bibnamefont
  {Li}}, \bibinfo {author} {\bibfnamefont {G.}~\bibnamefont {Wang}}, \ and\
  \bibinfo {author} {\bibfnamefont {Y.}~\bibnamefont {Wu}},\ }\href {\doibase
  10.1088/0954-3899/38/11/115004} {\bibfield  {journal} {\bibinfo  {journal}
  {J. Phys.}\ }\textbf {\bibinfo {volume} {G38}},\ \bibinfo {pages} {115004}
  (\bibinfo {year} {2011})}\BibitemShut {NoStop}%
\bibitem [{\citenamefont {Fu}\ \emph {et~al.}(2010)\citenamefont {Fu},
  \citenamefont {Liu},\ and\ \citenamefont {Wu}}]{Fu:2009wy}%
  \BibitemOpen
  \bibfield  {author} {\bibinfo {author} {\bibfnamefont {W.-j.}\ \bibnamefont
  {Fu}}, \bibinfo {author} {\bibfnamefont {Y.-x.}\ \bibnamefont {Liu}}, \ and\
  \bibinfo {author} {\bibfnamefont {Y.-L.}\ \bibnamefont {Wu}},\ }\href
  {\doibase 10.1103/PhysRevD.81.014028} {\bibfield  {journal} {\bibinfo
  {journal} {Phys. Rev.}\ }\textbf {\bibinfo {volume} {D81}},\ \bibinfo {pages}
  {014028} (\bibinfo {year} {2010})},\ \Eprint {http://arxiv.org/abs/0910.5783}
  {arXiv:0910.5783 [hep-ph]} \BibitemShut {NoStop}%
\bibitem [{\citenamefont {Cheng}\ \emph {et~al.}(2009)\citenamefont {Cheng}
  \emph {et~al.}}]{Cheng:2008zh}%
  \BibitemOpen
  \bibfield  {author} {\bibinfo {author} {\bibfnamefont {M.}~\bibnamefont
  {Cheng}} \emph {et~al.},\ }\href@noop {} {\bibfield  {journal} {\bibinfo
  {journal} {Phys. Rev.}\ }\textbf {\bibinfo {volume} {D79}},\ \bibinfo {pages}
  {074505} (\bibinfo {year} {2009})},\ \Eprint {http://arxiv.org/abs/0811.1006}
  {arXiv:0811.1006 [hep-lat]} \BibitemShut {NoStop}%
\bibitem [{\citenamefont {Aggarwal}\ \emph {et~al.}(2010)\citenamefont
  {Aggarwal} \emph {et~al.}}]{Aggarwal:2010wy}%
  \BibitemOpen
  \bibfield  {author} {\bibinfo {author} {\bibfnamefont {M.~M.}\ \bibnamefont
  {Aggarwal}} \emph {et~al.} (\bibinfo {collaboration} {STAR}),\ }\href
  {\doibase 10.1103/PhysRevLett.105.022302} {\bibfield  {journal} {\bibinfo
  {journal} {Phys. Rev. Lett.}\ }\textbf {\bibinfo {volume} {105}},\ \bibinfo
  {pages} {022302} (\bibinfo {year} {2010})},\ \Eprint
  {http://arxiv.org/abs/1004.4959} {arXiv:1004.4959 [nucl-ex]} \BibitemShut
  {NoStop}%
\bibitem [{\citenamefont {Adamczyk}\ \emph {et~al.}(2014)\citenamefont
  {Adamczyk} \emph {et~al.}}]{Adamczyk:2013dal}%
  \BibitemOpen
  \bibfield  {author} {\bibinfo {author} {\bibfnamefont {L.}~\bibnamefont
  {Adamczyk}} \emph {et~al.} (\bibinfo {collaboration} {STAR}),\ }\href
  {\doibase 10.1103/PhysRevLett.112.032302} {\bibfield  {journal} {\bibinfo
  {journal} {Phys. Rev. Lett.}\ }\textbf {\bibinfo {volume} {112}},\ \bibinfo
  {pages} {032302} (\bibinfo {year} {2014})},\ \Eprint
  {http://arxiv.org/abs/1309.5681} {arXiv:1309.5681 [nucl-ex]} \BibitemShut
  {NoStop}%
\bibitem [{\citenamefont {Rustamov}(2017)}]{Rustamov:2017lio}%
  \BibitemOpen
  \bibfield  {author} {\bibinfo {author} {\bibfnamefont {A.}~\bibnamefont
  {Rustamov}} (\bibinfo {collaboration} {ALICE}),\ }\bibfield  {booktitle}
  {\emph {\bibinfo {booktitle} {{Proceedings, 26th International Conference on
  Ultra-relativistic Nucleus-Nucleus Collisions (Quark Matter 2017): Chicago,
  Illinois, USA, February 5-11, 2017}}},\ }\href {\doibase
  10.1016/j.nuclphysa.2017.05.111} {\bibfield  {journal} {\bibinfo  {journal}
  {Nucl. Phys.}\ }\textbf {\bibinfo {volume} {A967}},\ \bibinfo {pages} {453}
  (\bibinfo {year} {2017})},\ \Eprint {http://arxiv.org/abs/1704.05329}
  {arXiv:1704.05329 [nucl-ex]} \BibitemShut {NoStop}%
\bibitem [{\citenamefont {Kitazawa}\ and\ \citenamefont
  {Asakawa}(2012{\natexlab{a}})}]{Kitazawa:2011wh}%
  \BibitemOpen
  \bibfield  {author} {\bibinfo {author} {\bibfnamefont {M.}~\bibnamefont
  {Kitazawa}}\ and\ \bibinfo {author} {\bibfnamefont {M.}~\bibnamefont
  {Asakawa}},\ }\href {\doibase 10.1103/PhysRevC.85.021901} {\bibfield
  {journal} {\bibinfo  {journal} {Phys. Rev.}\ }\textbf {\bibinfo {volume}
  {C85}},\ \bibinfo {pages} {021901} (\bibinfo {year} {2012}{\natexlab{a}})},\
  \Eprint {http://arxiv.org/abs/1107.2755} {arXiv:1107.2755 [nucl-th]}
  \BibitemShut {NoStop}%
\bibitem [{\citenamefont {Kitazawa}\ and\ \citenamefont
  {Asakawa}(2012{\natexlab{b}})}]{Kitazawa:2012at}%
  \BibitemOpen
  \bibfield  {author} {\bibinfo {author} {\bibfnamefont {M.}~\bibnamefont
  {Kitazawa}}\ and\ \bibinfo {author} {\bibfnamefont {M.}~\bibnamefont
  {Asakawa}},\ }\href {\doibase 10.1103/PhysRevC.86.024904,
  10.1103/PhysRevC.86.069902} {\bibfield  {journal} {\bibinfo  {journal} {Phys.
  Rev.}\ }\textbf {\bibinfo {volume} {C86}},\ \bibinfo {pages} {024904}
  (\bibinfo {year} {2012}{\natexlab{b}})},\ \bibinfo {note} {[Erratum: Phys.
  Rev. C86, 069902 (2012)]},\ \Eprint {http://arxiv.org/abs/1205.3292}
  {arXiv:1205.3292 [nucl-th]} \BibitemShut {NoStop}%
\bibitem [{\citenamefont {Bzdak}\ \emph {et~al.}(2018)\citenamefont {Bzdak},
  \citenamefont {Koch}, \citenamefont {Oliinychenko},\ and\ \citenamefont
  {Steinheimer}}]{Bzdak:2018uhv}%
  \BibitemOpen
  \bibfield  {author} {\bibinfo {author} {\bibfnamefont {A.}~\bibnamefont
  {Bzdak}}, \bibinfo {author} {\bibfnamefont {V.}~\bibnamefont {Koch}},
  \bibinfo {author} {\bibfnamefont {D.}~\bibnamefont {Oliinychenko}}, \ and\
  \bibinfo {author} {\bibfnamefont {J.}~\bibnamefont {Steinheimer}},\ }\href
  {\doibase 10.1103/PhysRevC.98.054901} {\bibfield  {journal} {\bibinfo
  {journal} {Phys. Rev.}\ }\textbf {\bibinfo {volume} {C98}},\ \bibinfo {pages}
  {054901} (\bibinfo {year} {2018})},\ \Eprint
  {http://arxiv.org/abs/1804.04463} {arXiv:1804.04463 [nucl-th]} \BibitemShut
  {NoStop}%
\bibitem [{\citenamefont {He}\ and\ \citenamefont {Luo}(2017)}]{He:2017zpg}%
  \BibitemOpen
  \bibfield  {author} {\bibinfo {author} {\bibfnamefont {S.}~\bibnamefont
  {He}}\ and\ \bibinfo {author} {\bibfnamefont {X.}~\bibnamefont {Luo}},\
  }\href {\doibase 10.1016/j.physletb.2017.10.030} {\bibfield  {journal}
  {\bibinfo  {journal} {Phys. Lett.}\ }\textbf {\bibinfo {volume} {B774}},\
  \bibinfo {pages} {623} (\bibinfo {year} {2017})},\ \Eprint
  {http://arxiv.org/abs/1704.00423} {arXiv:1704.00423 [nucl-ex]} \BibitemShut
  {NoStop}%
\bibitem [{\citenamefont {Bzdak}\ \emph {et~al.}(2017)\citenamefont {Bzdak},
  \citenamefont {Koch},\ and\ \citenamefont {Strodthoff}}]{Bzdak:2016sxg}%
  \BibitemOpen
  \bibfield  {author} {\bibinfo {author} {\bibfnamefont {A.}~\bibnamefont
  {Bzdak}}, \bibinfo {author} {\bibfnamefont {V.}~\bibnamefont {Koch}}, \ and\
  \bibinfo {author} {\bibfnamefont {N.}~\bibnamefont {Strodthoff}},\ }\href
  {\doibase 10.1103/PhysRevC.95.054906} {\bibfield  {journal} {\bibinfo
  {journal} {Phys. Rev.}\ }\textbf {\bibinfo {volume} {C95}},\ \bibinfo {pages}
  {054906} (\bibinfo {year} {2017})},\ \Eprint
  {http://arxiv.org/abs/1607.07375} {arXiv:1607.07375 [nucl-th]} \BibitemShut
  {NoStop}%
\bibitem [{\citenamefont {Luo}(2015{\natexlab{a}})}]{Luo:2015ewa}%
  \BibitemOpen
  \bibfield  {author} {\bibinfo {author} {\bibfnamefont {X.}~\bibnamefont
  {Luo}} (\bibinfo {collaboration} {STAR}),\ }\bibfield  {booktitle} {\emph
  {\bibinfo {booktitle} {{Proceedings, 9th International Workshop on Critical
  Point and Onset of Deconfinement (CPOD 2014): Bielefeld, Germany, November
  17-21, 2014}}},\ }\href@noop {} {\bibfield  {journal} {\bibinfo  {journal}
  {PoS}\ }\textbf {\bibinfo {volume} {CPOD2014}},\ \bibinfo {pages} {019}
  (\bibinfo {year} {2015}{\natexlab{a}})},\ \Eprint
  {http://arxiv.org/abs/1503.02558} {arXiv:1503.02558 [nucl-ex]} \BibitemShut
  {NoStop}%
\bibitem [{XL-(2018)}]{XL-private}%
  \BibitemOpen
  \href@noop {} {\bibfield  {journal} {\bibinfo  {journal} {X. Luo (STAR),
  private communication}\ } (\bibinfo {year} {2018})}\BibitemShut {NoStop}%
\bibitem [{\citenamefont {Davison}(2003)}]{davison2003statistical}%
  \BibitemOpen
  \bibfield  {author} {\bibinfo {author} {\bibfnamefont {A.}~\bibnamefont
  {Davison}},\ }\href {https://books.google.com/books?id=gQyIGGAiN4AC} {\emph
  {\bibinfo {title} {Statistical Models}}},\ Cambridge Series in Statistical
  and Probabilistic Mathematics\ (\bibinfo  {publisher} {Cambridge University
  Press},\ \bibinfo {year} {2003})\BibitemShut {NoStop}%
\bibitem [{\citenamefont {Luo}(2015{\natexlab{b}})}]{Luo:2014rea}%
  \BibitemOpen
  \bibfield  {author} {\bibinfo {author} {\bibfnamefont {X.}~\bibnamefont
  {Luo}},\ }\href {\doibase 10.1103/PhysRevC.91.034907} {\bibfield  {journal}
  {\bibinfo  {journal} {Phys. Rev.}\ }\textbf {\bibinfo {volume} {C91}},\
  \bibinfo {pages} {034907} (\bibinfo {year} {2015}{\natexlab{b}})},\ \Eprint
  {http://arxiv.org/abs/1410.3914} {arXiv:1410.3914 [physics.data-an]}
  \BibitemShut {NoStop}%
\bibitem [{sta(2014)}]{star_note}%
  \BibitemOpen
  \href@noop {} {\bibfield  {journal} {\bibinfo  {journal} {STAR Note 598,
  https://drupal.star.bnl.gov/STAR/ starnotes/public/sn0598}\ } (\bibinfo
  {year} {2014})}\BibitemShut {NoStop}%
\end{thebibliography}%

\end{document}